\documentclass[aps,manuscript]{revtex4}
\usepackage[T1]{fontenc}
\usepackage[latin9]{inputenc}
\usepackage{graphicx}

\makeatletter

\newcommand{\lyxdot}{.}

\@ifundefined{textcolor}{}
{%
 \definecolor{BLACK}{gray}{0}
 \definecolor{WHITE}{gray}{1}
 \definecolor{RED}{rgb}{1,0,0}
 \definecolor{GREEN}{rgb}{0,1,0}
 \definecolor{BLUE}{rgb}{0,0,1}
 \definecolor{CYAN}{cmyk}{1,0,0,0}
 \definecolor{MAGENTA}{cmyk}{0,1,0,0}
 \definecolor{YELLOW}{cmyk}{0,0,1,0}
}

\makeatother

\begin{document}

\title{Breakdown of effective phonon theory in one-dimensional chains with
asymmetric interactions}

\author{Yong Zhang}

\author{Shunda Chen}

\author{Jiao Wang}

\author{Hong Zhao}

\date{today}

\affiliation{Department of Physics and Institute of Theoretical Physics and Astrophysics,
Xiamen University, Xiamen 361005, Fujian, China}
\begin{abstract}
Phonons are universal language for solid state theory. Effective phonon
theory (self-consistent harmonic approximation) have been extensively
used to access the weakly nonlinear effect in solid materials, which
are standard content in textbook. In this manuscript, we test the
effective phonon theory in one-dimensional chains with asymmetric
interactions. It is found that simulation results agree well with
theoretical predictions for symmetric cases, while significantly deviate
from theoretical predictions for asymmetric cases. Our results imply
the asymmetric interaction can provide stronger phonon-phonon interaction,
which lead to break down of the effective phonon theory.
\end{abstract}
\maketitle
Nonlinearity play important roles in solid materials, which responds
for thermodynamic and transport properties such as thermal expansion
and thermal conductivity. Effective phonon theory (EPT) have been
commonly used to assess the \emph{weakly} nonlinear effect on phonons
(normal modes in classical physics), whose basic ideal is to find
a\emph{ effective harmonic} Hamiltonian to approximate the true nonlinear
Hamiltonian (in early literature, also called self-consistent harmonic
approximation) \cite{PhononPhys}. In the last decade, EPT have been
extensively used to study low dimensional lattice models. These studies
focus on one-dimensional (1D) momentum-conserved lattices and reported
that the numerical results agree well with theoretical predictions
even in strong nonlinearity limit \cite{AlabisoEP,BaowenEP,CaiEP,HeEP,LepriEP}.
It seemingly suggest EPT worked well for 1D lattice systems. However,
Our investigation here shows things are not that simple. We find that
simulation results agree well with theoretical predictions for 1D
chains with symmetric nonlinear interactions, while obviously deviate
from predictions for ones with asymmetric interactions. 

A 1D momentum-conserved lattice are generally defined by the dimensionless
Hamiltonian 
\begin{equation}
H=\sum_{i}\frac{p_{i}^{2}}{2}+V(x_{i}-x_{i-1})\label{eq:1}
\end{equation}
where $p_{i}$ and $x_{i}$ are the $i$th particle momentum and displacement
from the equilibrium position, respectively. $V$ is the potential
between two neighboring particles, which is given by 
\begin{equation}
V(x)=\frac{1}{2}x^{2}+\frac{\alpha}{3}x^{3}+\frac{\beta}{4}x^{4}\label{eq:2}
\end{equation}
The lattice spacing is set to unity. This means the sound velocity
$c_{s}$ in the harmonic limit is equal to one in our models. The
EPT calculate the effective particle force constants through self-consistently
replacing the harmonic force constants by their thermal averages over
all possible motions of particles. As a result, an effective harmonic
potential approximate the nonlinear one (2) by 
\begin{equation}
\widetilde{V}(x)=\frac{\eta^{2}}{2}x^{2}\label{eq:3}
\end{equation}
where $\eta$ is the renormalization factor. To do so, the phonon
frequencies are renormalized to $\widetilde{\omega}_{k}=\eta\omega_{k}$
by the factor $\eta$ as well, where $k$ is mode number and $\omega$
is the phonon frequency for (2) in harmonic limit. $\eta$ is usually
considered as $k$-independent, which is numerically confirmed in
FPU-$\beta$ models \cite{AlabisoEP,CaiEP}. Many theoretical methods
were used to deal with model (2) to calculate renormalized factor
$\eta$, such as generalized virial theorem \cite{AlabisoEP,BaowenEP},
Zwanzig-Mori projection \cite{LepriEP}, weak turbulence theory \cite{CaiEP}
and self-consistent phonon theory \cite{HeEP}. It have been confirmed
that these methods are equivalent \cite{CaiEP,HeEP,LepriEP}. In the
present manuscript, we test two two kind of expressions for $\eta$
based on the theories mentioned above. One is given by \cite{CaiEP}
\begin{equation}
\eta=\sqrt{\frac{<K>}{<U_{h}>}}\label{eq:4}
\end{equation}
where $K$ and $U_{h}$ is the total kinetic and harmonic potential
energy of model (2). Another slightly different expression was given
by \cite{BaowenEP}
\begin{equation}
\eta=\sqrt{\frac{<\sum_{i=1}^{N}\delta_{i}^{2}>+\alpha<\sum_{i=1}^{N}\delta_{i}^{3}>+\beta<\sum_{i=1}^{N}\delta_{i}^{4}>}{<\sum_{i=1}^{N}\delta_{i}^{2}>}}\label{eq:5}
\end{equation}
where $\delta_{i}\equiv x_{i}-x_{i-1}$. $<\cdots>$ in the above
two expressions denotes averaging over the Gibbs measure. 

We divide the potential (\ref{eq:2}) into two types according to
its symmetry with respect to its zero point at $x=0$. In the symmetric
type, we fix $\alpha=0$, which recovers the Fermi-Pasta-Ulam $\beta$
(FPU-$\beta$) models. In the asymmetric one, we fix $\beta=1$ and
varied $\alpha$ in the interval $0~2$ to ensure only one potential
minimum at $x=0$, which is the Fermi-Pasta-Ulam $\alpha\beta$ (FPU-$\alpha\beta$)
models. Nonlinear parameter $\alpha$ also governs the degree of the
interaction asymmetry. By increasing $\alpha$ from zero where the
potential is symmetric, one gets increasingly stronger asymmetry.

The effective sound velocity $\widetilde{c}_{s}$ is an important
outcome from EPT, which can be measured in a laboratory. Therefore,
it is natural to consider $\widetilde{c}_{s}$ as a observable to
verify the effectiveness of EPT. Obviously, $\widetilde{c}_{s}$ equals
the renormalization factor $\eta$ in our dimensionless models. On
the other hand, we can employ equilibrium fluctuation correlation
method \cite{EFCZhao} (for an more detailed version, see \cite{EFCZhao2})
to directly determine $\widetilde{c}_{s}$ by simulations. The method
has also been used to measure the effective sound velocity in FPU-$\beta$
models and the results agree well with the theoretical predictions
from Eqn. (5) \cite{BaowenEP2}. We conduct numerical simulations
at constant energy by integrating the equations of motion governed
by (2) with periodic boundary conditions. We use random initial conditions
by assigning velocities from a Gaussian distribution and zero relative
displacements for each particle, respectively, with the two constraints
that the total momentum is zero and the total energy is set to be
a specified constant. The Runge-Kutta algorithm of 7-8th order is
adopted with the time step $0.01$ to ensures the conservation of
the total system energy and momentum up to high accurate in the all
run time. The system then evolves for a certain transient time in
order to start the measurements in well-defined equilibrium states.
In order to confirm that the system has reached the thermal equilibrium
state, we have checked that the value of energy localization to be
$O(1)$ and the relative displacement and the velocity distributions
of particles agrees well with the Gibbs statistics \cite{CaiEP}.

In Fig. 1, we compare the measured $\widetilde{c}_{s}$ as a function
of nonlinear parameters with the theoretical predictions from Eqs.
(4) and (5) where $<\cdots>$ is replaced by time average. It is clearly
seen that the measured results agree well with the predictions of
EPT for FPU-$\beta$ chains in the entire range of $\beta$ under
study. In contrast, the systematic deviation from the predictions
is clearly revealed as $\alpha$ increasing for FPU-$\alpha\beta$
chains. The theoretical results obviously underestimate the effective
sound velocity of FPU-$\alpha\beta$ chains with larger asymmetry.
These results imply that EPT does well for nonlinear chins with symmetric
interactions, but break down for ones with asymmetric interactions.

\begin{figure}
\includegraphics[scale=0.5]{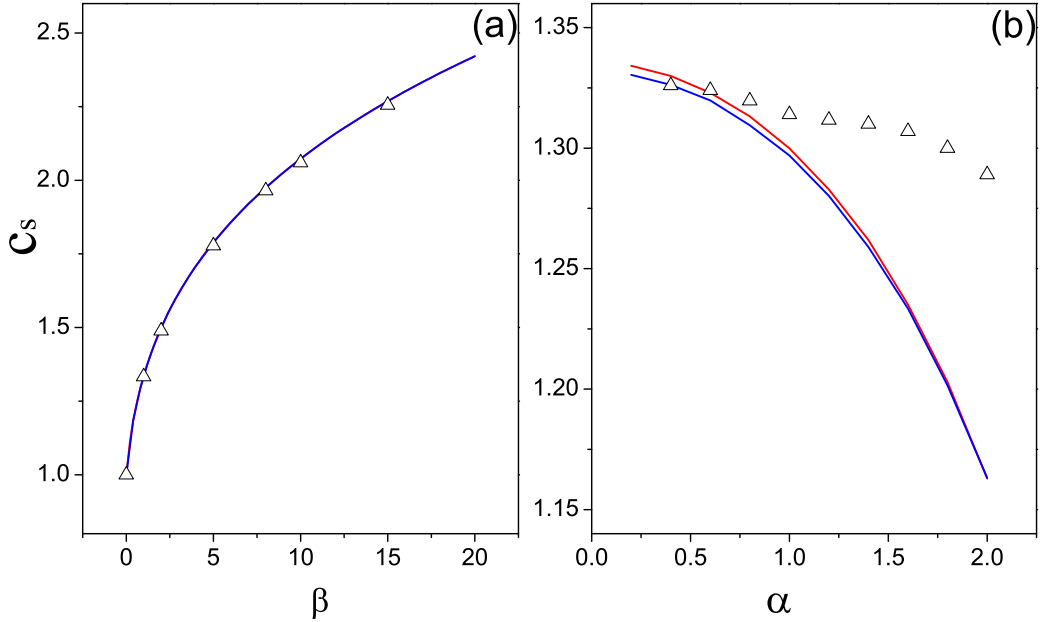}

\caption{The effective sound velocity is plotted versus nonlinear parameters
$\beta$ in FPU-$\beta$ chains (a) and $\alpha$ in FPU-$\alpha\beta$
chains (b) at equilibrium temperatures $T=0.5$. The open triangles
indicate the measured value by the equilibrium fluctuation correlation
method for $N=2048$ \cite{EFCZhao}. The red lines indicate the predictions
from the Eqn. (4) and the blue lines indicate the predictions from
the Eqn. (5), where $N=128$.}

\end{figure}

For understanding the breakdown of EPT in 1D chains with asymmetric
interactions, we shall perform a more microscopic measurement as follows.
We directly extract the phonon frequency $\omega_{m}$ from the true
dynamics governed by (2), then calculate the measured renormalization
factor by $\eta_{m}=\frac{\omega_{m}}{\omega_{h}}$, where $\omega_{h}$
is the corresponding harmonic phonon frequency. To this aim, we shall
analyze the power spectrum of velocity of a singe particle randomly
picked up from the chain at equilibrium. The same simulation process
as mentioned above is used. A FFT algorithm is used to obtain the
power spectrum with a time series of total time $10^{5}$ to ensure
the high frequency resolution. The power spectrum is averaging over
at least 20 trajectories to suppress the statistical fluctuations.

Figure. 2 shows the power spectrum of the single-particle velocity
time series for FPU-$\beta$ model with $\beta$=10 and FPU-$\alpha\beta$
model with $\alpha$=2. In spite of strong nonlinearity and long system
size ($N=2048$ for both models), it is clearly seen that a few phonon
peaks with the lowest frequencies is well distinguished from others
whose peaks quickly decay into noise signals due to stronger the phonon-phonon
interactions. It was commonly accepted that the long-wavelength phonons
dominate the asymptotic behavior of 1D momentum-conserved lattices.
This is the very reason that a long-wavelength approximations was
extensively adopted in varied theories, such as the self-consistent
mode-coupling theory \cite{modecoupling} and hydrodynamics \cite{hydrodynamics1,hydrodynamics2,Spohn}.
In addition, some important physical quantity, such as sound velocity,
is defined in a long-wavelength limit. Therefore, it is fair to focus
on \emph{the lowest frequency phonon peak}. The separability of the
lowest frequency phonon peak have been checked for all parameters
under investigation, and hence allow us to accurately measure its
frequency and renormalization factor. Interestingly, our results intuitively
display the image of ``effective phonons'' that was proposed on
purely phenomenological basis in varied theories. 

\begin{figure}
\includegraphics[scale=0.5]{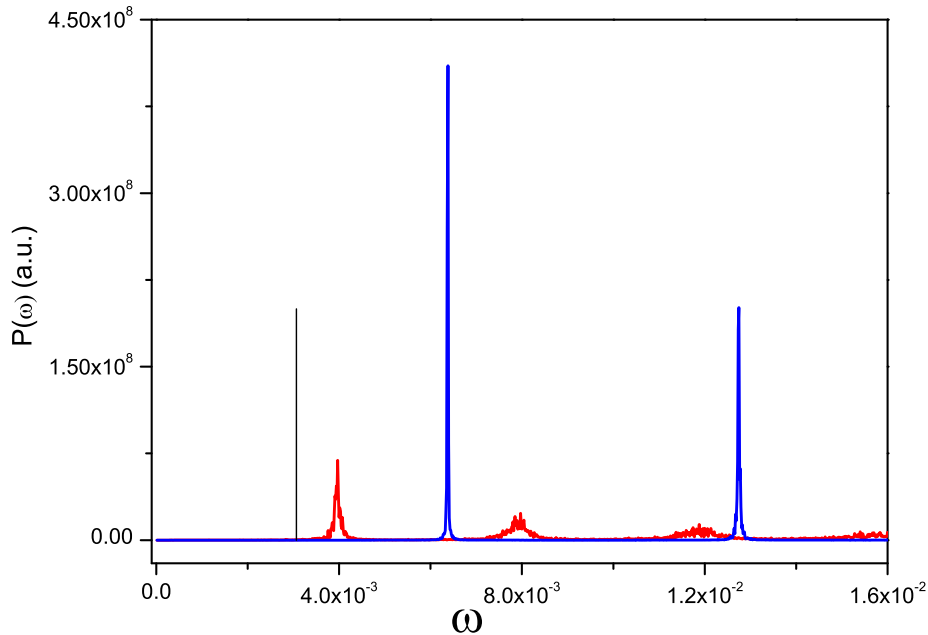}

\caption{The phonon peaks in the low frequency regim for $N=2048$and $T=0.5.$Blue
solid line indicate the FPU-$\beta$ with $\beta=10$; Red solid line
indicate the FPU-$\alpha\beta$ with $\alpha=2$ and $\beta=1$. For
comparison, the lowest harmonic frequency is also plotted by black
solid line.}

\end{figure}

Figure. 2 also shows a notable difference between FPU-$\alpha\beta$
and FPU-$\beta$ models for linewidths of phonon peaks. The phonon
peak for FPU-$\alpha\beta$ model is broader than the one for FPU-$\beta$
models by one order of magnitude under the current parameters. That
means the asymmetric interactions provide stronger phonon-phonon interaction
compared to the symmetric ones, which lead to shorter phonon lifetime.
The results on phonon linewidths will be reported in our otherwise
manuscript. Here, we focus on the effect of nonlinearity on phonon
frequency. 

In Fig. 3, we plot the measured renormalization factor $\eta_{m}$
as the function of nonlinear parameters for two representative equilibrium
temperatures $0.1$ and $0.5$, respectively. Meanwhile, we calculate
the theoretical value of $\eta$ via the expressions (4) and (5),
in which $<\cdots>$ was replaced by time average. For FPU-$\beta$
chains, the measured factor $\eta_{m}$ agree well with theoretical
predictions in the entire parameter region being explored at both
temperatures. In sharp contrast to that, the systematic deviation
from the prediction is clearly seen again for FPU-$\alpha\beta$ chains
at larger $\alpha$ at both temperatures. We also check that the $\eta_{m}$
is independent of systems size $N$ for both FPU-$\beta$ and FPU-$\alpha\beta$
models, as shown in Fig. 4. This result implies that the measured
$\eta_{m}$ is independent of the wave numbers of phonons for the
two models. Therefore, It is reasonable to measure the renormalization
factor via the single phonon peak with the lowest frequency. The results
further reveal that the EPT fail to renormalize the phonon frequency
in chains with asymmetric interactions. 

\begin{figure}
\includegraphics[scale=0.5]{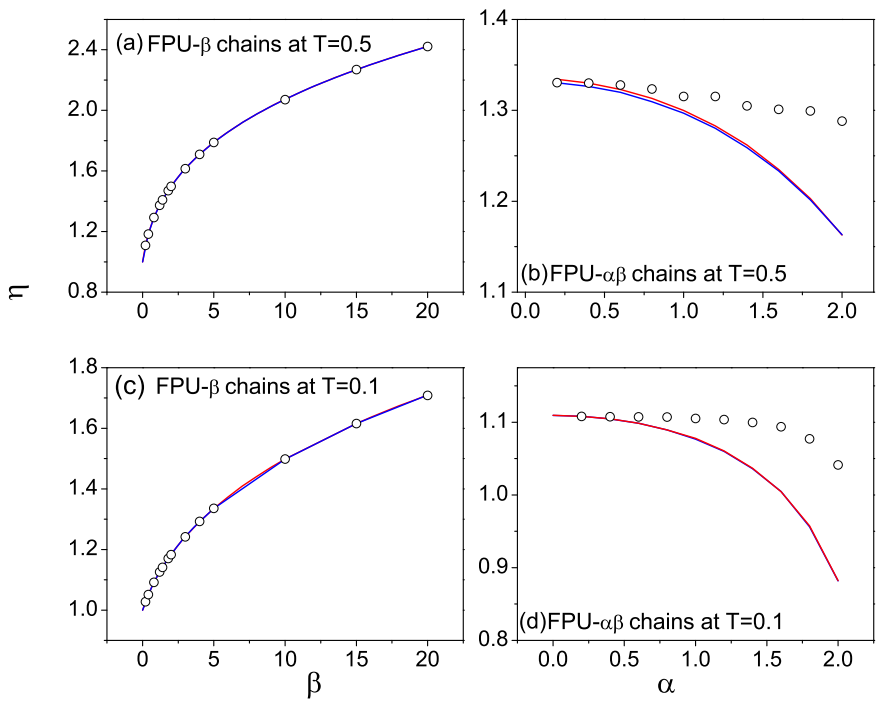}

\caption{The renormalization factor is plotted versus nonlinear parameters
$\beta$ in FPU-$\beta$ chains and $\alpha$ in FPU-$\alpha\beta$
chains at equilibrium temperatures $T=0.5$ and $T=0.1$. The open
circles indicate the measured value $\eta_{m}$. The red lines indicate
the predictions from the Eqn. (4) and the blue lines indicate the
predictions from the Eqn. (5). }

\end{figure}

Naturally, a following question is if the measured renormalization
factor can exactly predict the effective sound velocity for the chains
under investigation. To check this, we compare the effective sound
velocity obtained by $\eta$ with the measured one from the equilibrium
fluctuation correlation method mentioned above. In Fig. 5, it is clearly
seen that there is an excellent agreement between these two ways to
obtain the effective sound velocity in both FPU-$\beta$ abnd FPU-$\alpha\beta$
models. These results confirm that one can obtain the correct renormalized
frequency of phonon even in chains with larger interaction asymmetry
where the EPT break down.

\begin{figure}
\includegraphics[scale=0.5]{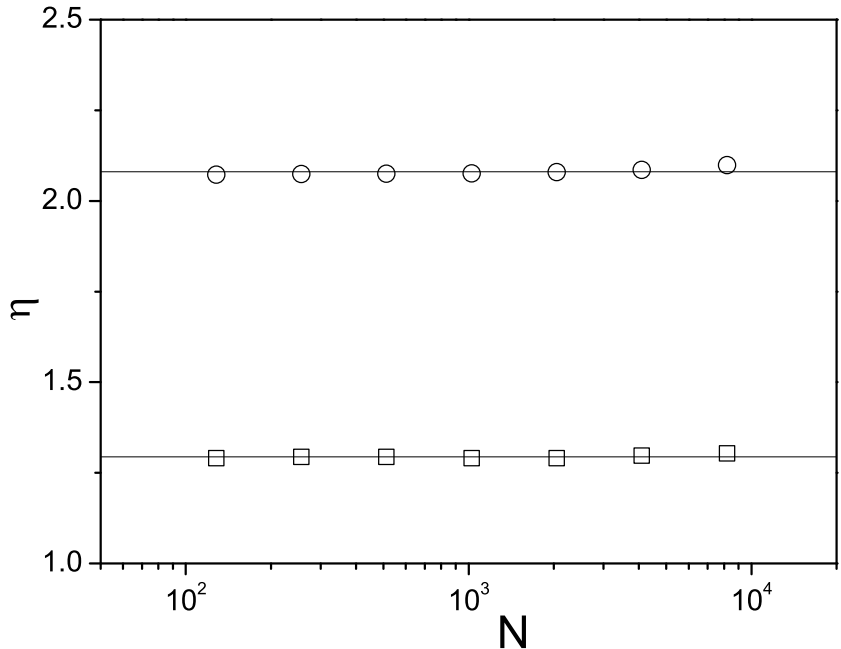}

\caption{The measured renormalization factor is plotted versus system size
$N$ at equilibrium temperature $T=0.5$ for FPU-$\beta$ chains with
$\beta=10$ (open circles) and FPU-$\alpha\beta$ chains with $\alpha$=2
and $\beta=1$(open squares), where two horizontal lines are responding
to their mean values, respectively.}

\end{figure}

The effectiveness of EPT for FPU-$\beta$ chains even in a strongly
nonlinear regime has also been confirmed by other authors \cite{AlabisoEP,BaowenEP,CaiEP},
and be well understood under the frame of weak turbulence theory \cite{CaiEP}.
In these theoretical scenario, the FPU-$\beta$ dynamics with the
strong nonlinear limit in thermal equilibrium can be effectively described
by a system of weakly interacting renormalized normal modes. Such
effective renormalization results mainly from the trivial resonant
wave interactions, i.e., interactions with no momentum exchange. It
is thus approprite that fro FPU-$\beta$ chains the mean-field approximation
is employed as does EPT.

\begin{figure}
\includegraphics[scale=0.5]{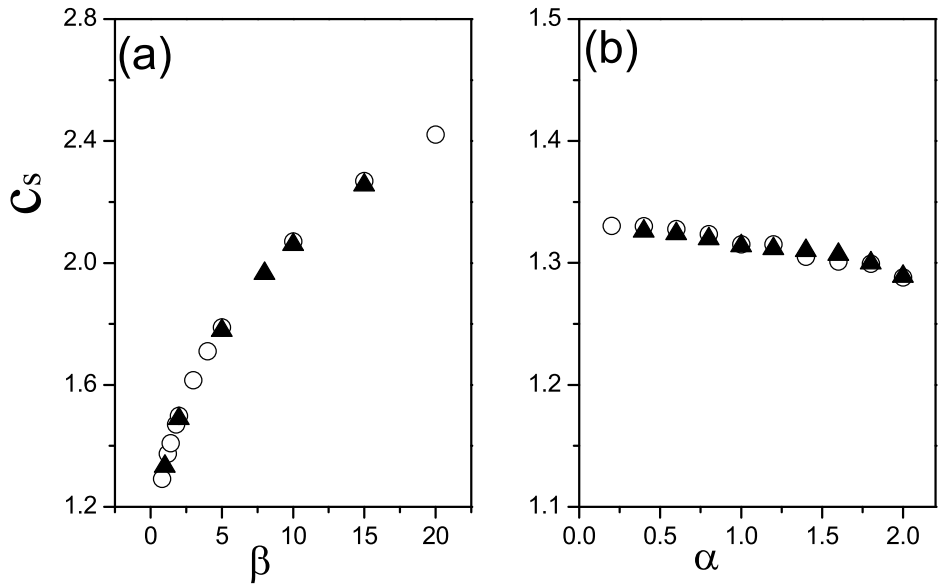}

\caption{The measured sound velocity is plotted versus nonlinear parameters
$\beta$ for FPU-$\beta$ chains (a) and FPU-$\alpha\beta$ chains
(b) at the equilibrium temperature $T=0.5$. The open circles indicate
the values obtained by $\eta_{m}$based on a analysis of the lowest
frequency phonon peak for $N=128$. The solid triangles indicate the
values from measured values from the equilibrium fluctuation correlation
method for $N=2048$ \cite{EFCZhao}.}

\end{figure}

However, such a picture seemingly can not extend to the chains with
asymmetric inter-particle interactions, as revealed by the results
mentioned above. The deviation from the calculations in EPT in FPU-$\alpha\beta$
chains suggest that the asymmetric interactions may provide stronger
nonlinearity than symmetric ones. Mulch-phonon processes beyond the
trivial resonant wave interactions should be considered, lead to breakdown
of EPT. In fact, it was also found that asymmetric interactions results
in stronger phonon-phonon interactions in the recent studies of energy
equipartition \cite{Eequi} and heat conduction \cite{symheat1,symheat2}
in FPU-$\alpha\beta$ chains. In the former case, shorter equipartition
time was observed for FPU-$\alpha\beta$ chains comparing to the FPU-$\beta$
chains, and in the latter case, normal heat conduction was reported,
against the abnormal heat conduction for FPU-$\beta$ chains. 

In summary, we check the EPT for bath FPU-$\beta$ and FPU-$\alpha\beta$
models through an analysis of the lowest frequency phonon peak. It
is found that the numerically measured renormalization factor and
sound velocity agree well with the theoretical predictions for FPU-$\beta$
chains, whereas obviously deviate from predictions for FPU-$\alpha\beta$
chains. We stress that the asymmetric interactions play an important
role responsible for breakdown of EPT, which provide stronger interactions
among the phonons beyond the ones without momentum exchange occurred
in the chains with symmetric interactions. We also confirm that the
analysis based on the lowest frequency phonon peak can obtain the
renormalized phonon frequency in momentum conserved chain with asymmetric
interactions where EPT break down. It is a challenge to correct theories
to effectively renormalize anharmonicity induced by asymmetric interactions.

Furthermore, the lowest frequency phonon peak analysis provide a test
bed for the theories based on the long-wavelength approximations,
such as the self-consistent mode coupling theory \cite{modecoupling}
and hydrodynamics \cite{hydrodynamics1,hydrodynamics2,Spohn}. The
current work together with our recent reports on heat conduction \cite{symheat1,symheat2}
have revealed the important roles of asymmetric interactions for phonon
interactions in 1D momentum conserved nonlinear chains, which have
been underestimated in conventional theories. 
\begin{acknowledgments}
!\end{acknowledgments}


\begin{thebibliography}{10}
\bibitem{PhononPhys}G. P. Srivastave, The Physics of Phonons (IOP,
Bristol, 1990).

\bibitem{AlabisoEP}C. Alabiso, M. Casrtelli, and P. Marenzoni, J.
Stat. Phys. 79, 451 (1995).

\bibitem{BaowenEP}Nianbei Li, Peiqing Tong, and Baowen Li, Europhys.
Lett. 75, 49 (2006).

\bibitem{LepriEP}S. Lepri, Phys. Rev. E 58, 7165 (1998).

\bibitem{CaiEP}B. Gershgorin, Yuri V. Lvov, and D. Cai, Phys. Rev.
Lett. 95, 264302 (2005); Phys. Rev. E 75, 046603 (2007).

\bibitem{HeEP}D. He, S. Buyukdagli, and B. Hu, Phys. Rev. E 78, 061103
(2008).

\bibitem{EFCZhao}H. Zhao, Phys. Rev. Lett. 96, 140602 (2006).

\bibitem{EFCZhao2}S. Chen, Y. Zhang, J. Wang, H. Zhao, arXiv:1106.2896

\bibitem{BaowenEP2}N. Li, B. Li, and S. Flach, Phys. Rev. Lett. 105,
054102 (2010).

\bibitem{modecoupling}L. Delfini, S. Lepri, R. Livi, and A. Politi,
Phys. Rev. E 73, 060201(R) (2006); J. Stat. Mech. (2007) P02007.

\bibitem{simpleliquids}J. P. Hansen and I. R. McDonald, Theory of
Simple Liquids, 3rd ed. (Academic, London, 2006).

\bibitem{hydrodynamics1}O. Narayan and S. Ramaswamy, Phys. Rev. Lett.
89, 200601 (2002).

\bibitem{Spohn}M. Prahofer and H. Spohn, J. Stat. Phys. 115, 255
(2004).

\bibitem{hydrodynamics2}H. van Beijeren, Phys. Rev. Lett. 108 180601
(2012).

\bibitem{Eequi}G. Benettin and A. Ponno, J. Stat. Phys. 144, 793
(2011).

\bibitem{symheat1}Yi Zhong, Yong Zhang, Jiao Wang, and Hong Zhao,
Phys. Rev. E 85, 060102(R) (2012)

\bibitem{symheat2}Shunda Chen, Yong Zhang, Jiao Wang, and Hong Zhao,
arXiv:1204.5933\end{thebibliography}
\end{document}